\documentclass[a4paper,preprint,superscriptaddress,preprintnumbers,nofootinbib]{revtex4}
\newcommand{\Slash}[1]{{\ooalign{\hfil/\hfil\crcr$#1$}}}
\newcommand{\dcsb}{D$\chi$SB}

\usepackage{amsmath}	
\usepackage[dvipdf]{graphicx}
\usepackage{bm}

\begin{document}

\title{Weak solution method of the non-perturbative renormalization group equation to describe dynamical chiral symmetry breaking and its application to beyond the ladder analysis in QCD
\footnote{Presented at ``SCGT12 KMI-GCOE Workshop on Strong Coupling Gauge
Theories in the LHC Perspective'', 4-7 Dec. 2012, Nagoya University.}}


\author{Ken-Ichi \surname{Aoki}}
\email{aoki@hep.s.kanazawa-u.ac.jp}
\affiliation{Institute for Theoretical Physics, Kanazawa University, Kanazawa 920-1192, Japan}

\author{Daisuke \surname{Sato}}
\email{satodai@hep.s.kanazawa-u.ac.jp}
\affiliation{Institute for Theoretical Physics, Kanazawa University, Kanazawa 920-1192, Japan}

\preprint{KANAZAWA-13-04}


\begin{abstract}
The method of non-perturbative renormalization group (NPRG) is applied to the analysis of dynamical chiral symmetry breaking (\dcsb) in QCD.
We show that the {\dcsb} solution of the NPRG flow equation can be obtained without the bosonization.
The solution, having the singular point, can be authorized as the weak solution of partial differential equation,
and can be easily evaluated using the method of the characteristic curve.
Also we show that our non-ladder extended approximation improves 
almost perfectly the gauge dependence of the chiral condensates.
\end{abstract}
\maketitle



\section{Introduction}
The approach of the ``non-perturbative renormalization group'' (NPRG) based on Wilson's idea allows us to analyze dynamical chiral symmetry breaking (\dcsb) in QCD with systematic approximations compared to the ladder Schwinger-Dyson analysis which suffers from the strong gauge dependence. 
The NPRG approach has been formulated as several forms of the functional differential equation, so people often call these formulations the ``functional renormalization group''.
Among them, we adopt the flow equation derived by Wetterich \cite{Wetterich:1992yh}.

In NPRG, the 4-fermi coupling constant dynamically generated by the gauge interaction blows up at an infrared scale as a signal of {\dcsb}. 
Due to its explosive behavior, the access to the broken symmetry scale below the critical scale is a nontrivial problem.
In the earlier studies, the auxiliary field (bosonization)\cite{Aoki96,Aoki:1999dw}, the dynamical bosonization\cite{Gies:2001nw}, and the explicit breaking term such as the bare mass term \cite{Aoki:2009zza} have been introduced to access the broken symmetry scale.
However, using the weak solution method\footnote{The authors greatly appreciate helpful comments by Prof. Akitaka Matsumura who told us how to construct the weak solution.}, we can  directly access the broken symmetry scale only with the multi-fermi interactions but without any help by those methods.


\section{Truncated Effective Action}
In order to study the dynamical chiral symmetry breaking in QCD with three-flavor massless quarks,
we define the following truncated effective action:
\begin{align}
\begin{split}
 \Gamma_\Lambda[\Phi]=&\int_x \bigg\{
 \frac{Z_A}{4}F^a_{\mu\nu}F_a^{\mu\nu}
 +\frac{1}{2\xi} \left(\partial_\mu A_\mu \right)^2
 +\bar{\psi}\left(Z_\psi \Slash{\partial}+i g_s\Slash{A}\right)\psi
 -V(\psi,\bar{\psi};\Lambda)
 \bigg\},\label{eq-app_eff_action}
\end{split}
\end{align}
where we use the covariant gauge with the gauge-fixing parameter $\xi$, and do not represent the ghost sector for simplicity.
In this definition, the multi-fermi interactions are represented by $V(\psi,\bar{\psi};\Lambda)$, which we call the fermion potential.

We substitute the effective action (\ref{eq-app_eff_action}) into the Wetterich flow equation, and expand it in terms of the fields and the derivatives, so that we can obtain the coupled non-perturbative RG equations of these coupling constants.
Then, lowering the cutoff scale $\Lambda$ by solving the RG equations, the gauge interaction dynamically induces multi-fermi interactions. Among them, the scalar 4-fermi interaction brings about the {\dcsb} at an intermediate cutoff scale as the Nambu--Jona-Lasinio model does.
The 4-fermi coupling constant blows up at a finite infrared scale $\Lambda_{\rm c}$,  showing the signal of the {\dcsb}, and consequently its RG flow cannot go beyond the critical scale $\Lambda_{\rm c}$.

\section{Partial Differential Equation}
\begin{figure}
\centering
\includegraphics[width=3.5in]{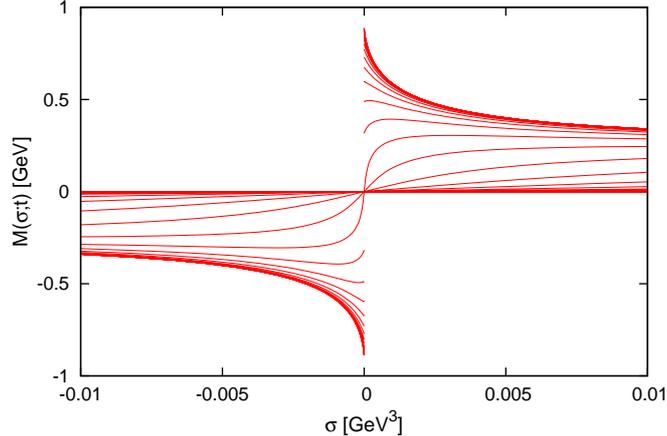}
\caption{ RG evolution of the mass function $M(\sigma;t)$.}
\label{fig-m-func_vs_sig}
\end{figure}

To go beyond the critical scale $\Lambda_{\rm c}$, we solve the flow equation without expanding the fermion potential with respect to the scalar fermion-bilinear field $\sigma(\equiv \bar{\psi}\psi)$ that is the central operator for {\dcsb}.
For simplicity, the arguments of the fermion potential are limited to the scalar operator: $V(\psi,\bar{\psi};\Lambda)\rightarrow V(\sigma;\Lambda)$.
Besides, the interactions are limited to the ladder type ones,
and consequently we obtain the ladder flow equation as the following partial differential equation (PDE)\cite{Aoki96, Aoki:1999dw}:
\begin{align}
 \partial_t V(\sigma;t) &=
 -\eta_\psi \sigma\partial_\sigma V+
\frac{\Lambda^4}{4\pi^2}\ln\left[1+
   \Lambda^{-2}\left(
      \partial_\sigma V+(3+\xi)\frac{ C_2\, g^2_{\rm s}\sigma}{4\Lambda^2}
               \right)^2
   \right],\label{eq-ladder-flow-equation}
\end{align}
where $C_2$ is the second Casimir invariant of $SU(3)$. 
Here the dimensionless scale parameter $t(\equiv \log\Lambda_0/\Lambda)$ and the anomalous dimension of the quark field $\eta_\psi(\equiv \partial_t \log Z_\psi)$ are introduced.
The PDE (\ref{eq-ladder-flow-equation}) can be numerically solved by replacing the derivatives with the finite differences. 
Practically, we work with the PDE for the first derivative of the fermion potential,  $M(\sigma;t)(\equiv \partial_\sigma V)$, which we call the mass function.
As shown in Fig.~\ref{fig-m-func_vs_sig}, the numerical solution of the mass function at $\sigma=0$ has a finite jump below the critical scale $\Lambda_{\rm c}$.
This is nothing but the signal of {\dcsb}, and the infrared value of the mass function at the limit, $\sigma\rightarrow +0 $, is nothing but the dynamical mass $m_{\rm dyn.}$

\section{Weak Solution}
While the singular behavior have numerically been evaluated, the non-analytic solution cannot be mathematically authorized as a solution of the PDE.
Such a solution can be defined  as the ``weak solution'' which satisfies the integral form of the PDE\cite{Eva01}.
Here we skip the detail of its definition, but briefly introduce the method of the characteristic curve to evaluate the weak solution.

\begin{figure}
\centering
\includegraphics[width=5.0in]{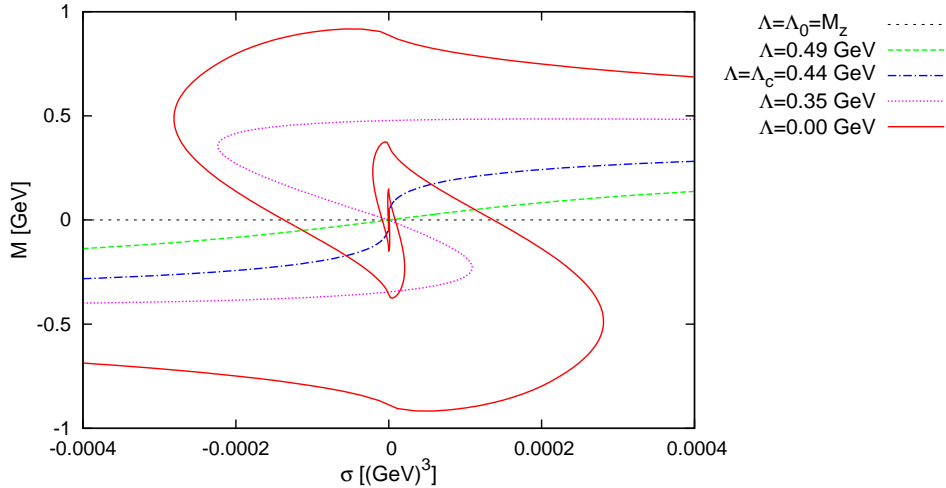}
\caption{Parametric curve of the mass function $M(\bar{\sigma};t)$ in terms of $\sigma_0$.}
\label{aba:fig1}
\end{figure}

Representing the right-hand side of Eq.~(\ref{eq-ladder-flow-equation}) as a function $F(\sigma, M;t)$,
the PDE for the mass function can be written by  
\begin{align}
 \partial_t M(\sigma;t)&=-F_\sigma - \partial_\sigma M\cdot F_M,\label{eq-pde_mass_func}
\end{align}
where we use the notations, $F_\sigma\equiv\partial_\sigma F(\sigma,M;t)|_M$, $F_M\equiv\partial_M F(\sigma,M;t)|_\sigma$.
Now let us consider the mass function on a characteristic curve, $\sigma=\bar{\sigma}(t;\sigma_0)$, where $\sigma_0$ is the initial position of $\bar{\sigma}$.
If $\bar{\sigma}(t;\sigma_0)$ obeys the following differential equation,
\begin{align}
 \frac{d}{dt} \bar{\sigma}(t;\sigma_0)&= F_M,\label{095929_28Mar13}
\end{align}
then the derivative $\partial_\sigma M$ vanishes in the PDE:
\begin{align}
 \frac{d}{dt} M(\bar{\sigma};t)& = -F_\sigma.\label{095938_28Mar13} 
\end{align}
Eqs.~(\ref{095929_28Mar13}) and (\ref{095938_28Mar13}) are coupled ordinary differential equations equivalent to the PDE (\ref{eq-pde_mass_func}).
The parametric curve of the mass function $M(\bar{\sigma};t)$ in terms of $\sigma_0$ can be constructed by solving them for each $\sigma_0$ as shown in Fig.~\ref{aba:fig1}.
Below the critical scale $\Lambda_{\rm c}$, the parametric curve is no longer regarded as the one-to-one mass function of $\sigma$.
However, a condition of the weak solution called by ``Rankine-Hugoniot condition'' uniquely gives the one-to-one function shown in Fig.~\ref{fig-m-func_vs_sig}\cite{Aoki:2013}.

\section{Result and Discussion}
Finally we show the result of the non-ladder extended flow equation, where the non-ladder corrections are added to the ladder flow equation (\ref{eq-ladder-flow-equation})\cite{Aoki:2012mj}.
In this approximation, the commutator of the generator of $SU(3)_{\rm c}$ is ignored consistently for obtaining the gauge independent result \cite{Aoki:2000dh}.

\begin{figure}
\centering
\includegraphics[width=3.5in]{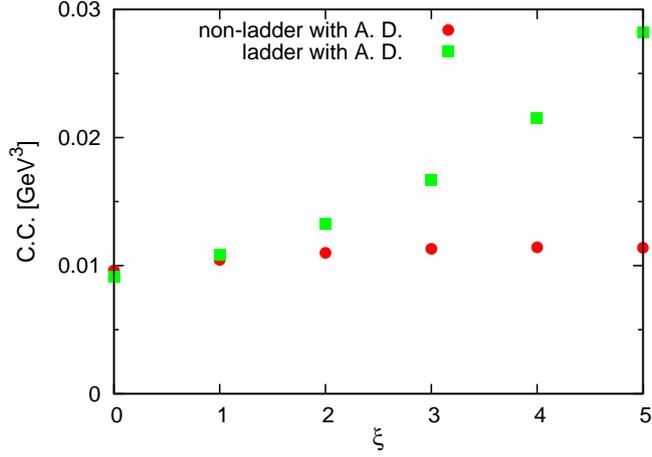}
\caption{The dependence of the chiral condensates $\langle\bar{\psi}\psi\rangle$ on the gauge-fixing parameter $\xi$.}
\label{fig-dependence_chi}
\end{figure}

The gauge dependence of the chiral condensates $\langle\bar{\psi}\psi\rangle$, which is actually obtained by introducing the bare mass term of quarks as its source term, is shown in Fig.~\ref{fig-dependence_chi}.
The gauge dependence of the non-ladder extended results is greatly suppressed compared with the ladder results.
Therefore we conclude that our non-ladder extended flow equation well respects the gauge independence.
Here it should be noted that, in the Landau gauge ($\xi=0$), the gauge-dependent ladder result of the chiral condensates coincides with the almost gauge-independent non-ladder extended one.
This feature of the Landau gauge proves a folklore that the ladder approximation looks good particularly in the Landau gauge.

\begin{thebibliography}{9}

\bibitem{Wetterich:1992yh}
  C.~Wetterich,
  {\em Phys.\ Lett.\ B} {\bf 301}, 90 (1993).

\bibitem{Aoki:2012mj} 
  K-I.~Aoki and D.~Sato,
 {\em Prog. Theor. Exp. Phys.} {\bf  2013}, 043B04 (2013). 



\bibitem{Aoki96} K-I.~Aoki, {\em Proc. SCGT96}, 171 (1996):hep-ph/9706264,
{\em Prog. Theor. Phys. Suppl.} {\bf 131}, 129 (1998), 
{\em Int. J. Mod. Phys. B} {\bf 14}, 1249 (2000).

\bibitem{Aoki:1999dw}
  K-I.~Aoki, K.~Morikawa, J.~-I.~Sumi, H.~Terao and M.~Tomoyose,
{\em Prog. Theor. Phys.} {\bf 102}, 1151 (1999), 
{\em Phys.\ Rev.\ D} {\bf 61}, 045008 (2000) .


 


\bibitem{Gies:2001nw}
  H.~Gies and C.~Wetterich,
  Phys.\ Rev.\ D {\bf 65}, 065001 (2002) .

\bibitem{Aoki:2009zza}
  K-I.~Aoki and K.~Miyashita,
  {\em Prog.\ Theor.\ Phys.}\  {\bf 121}, 875 (2009).

\bibitem{Eva01} L.~C.~Evans, {\em Partial Differential Equations}, 2nd ed. (AMS, 2010).


\bibitem{Aoki:2013}
 K-I.~Aoki, D.~Sato and S.-I.~Kumamoto, in preparation (2013).

\bibitem{Aoki:2000dh}
  K-I.~Aoki, K.~Takagi, H.~Terao and M.~Tomoyose,
  {\em Prog.\ Theor.\ Phys.}\  {\bf 103}, 815 (2000).

\end{thebibliography}
\end{document}